\documentclass[useAMS,usenatbib]{mn2e}

\usepackage{natbib}
\usepackage{graphicx}
\usepackage{txfonts}
\usepackage{mathrsfs}

\title[The Influence of Dust Formation Modelling  on Na~I and K~I Line Profiles  in Substellar Atmospheres]{The Influence of Dust Formation Modelling  on Na~I and K~I Line Profiles  in Substellar Atmospheres}
\author[C.~M.~S.\ Johnas, Ch.~Helling, M. Dehn, P.~Woitke, P.~H.\ Hauschildt]
       {C.~M.~S.\ Johnas$^{1}$\thanks{E-mail: yeti@hs.uni-hamburg.de},  
        Ch.~Helling$^{2}$,  M. Dehn$^{1}$,  P.~Woitke$^{3}$, P.~H.\ Hauschildt$^{1}$\\
$^{1}$Hamburger Sternwarte, Gojenbergsweg 112, 21029 Hamburg, Germany\\
$^{2}$ SUPA, School of Physics \& Astronomy, Univ. of St Andrews, North Haugh, St Andrews,  KY16 9SS, UK\\
$^{3}$ SUPA, Astronomy Technology Center, The University of Edinburgh, Royal Observatory, Blackford Hill, Edinburgh EH9 3HJ, Scotland, UK}

\begin{document}

\date{Accepted 2008 January 21.  Received 2008 January 21; in original form 2007 December 6}

\pagerange{\pageref{firstpage}--\pageref{lastpage}} \pubyear{2007}

\maketitle

\label{firstpage}

\begin{abstract}
  We aim to understand the correlation between cloud formation and
  alkali line formation in substellar atmospheres.We perform line
  profile calculations for Na~I and K~I based on the coupling of our
  kinetic model for the formation and composition of dust grains with
  1D radiative transfer calculations in atmosphere models for brown
  dwarfs and giant gas planets.  The Na~I and K~I line profiles
  sensibly depend on the way clouds are treated in substellar
  atmosphere simulations. The kinetic dust formation model results in
  the highest pseudo-continuum compared to the limiting cases.
\end{abstract}


\section{Introduction}

Dust influences the environment from which it forms by consuming
elements from the gas phase. Hence, dust selectively alters the local
metallicity in the atmosphere of a brown dwarf as well as in planetary
atmospheres (Woitke \& Helling 2004, Helling, Woitke \& Thi 2008).  Dust
clouds furthermore efficiently absorb photons which can
result in $\mu$m-broad absorption features (Helling et al. 2006, 2007; Cushing et
al. 2006, Burgasser et al. 2007). The absorbed energy is redistributed
inside the lattice structure of the solid grain causing an isotropic
irradiation of IR photons into the atmosphere, a process, which
considerably alters the atmospheric temperature profile (Tsuji et
al. 1996, Tsuji 2005, Ackerman \& Marley 2001, Allard et al. 2001).
Both processes, the selective alteration of local element abundances
and the changing atmospheric temperature, have a distinct influence on
the gas-phase chemistry which determines the atomic and molecular
concentrations. These intricate effects are important even for
elements not or only marginally involved into the dust formation process
as we will show in this letter.

Alkali lines are dominant features in the optical spectrum of brown
dwarfs and the far line wings of the most abundant alkali metals,
sodium and potassium, determine the pseudo-continuum in this spectral
range (Johnas et al. 2007a, Allard et al. 2007).  Another alkali, Li,
is used to determine fundamental parameter of substellar objects
(e.g. Martin, Rebolo \& Maguzza 1994, Johnas et al. 2007c).  They all
can be a powerful tool to study the cloud formation in substellar
objects if combined with an atmosphere simulation which allows to
correlate the line profiles with an analysis of the atmosphere
regarding the $(T, p_{\rm gas}, v_{\rm conv})$ structure ( $T$ - gas
temperature, $p_{\rm gas}$ - gas pressure, $v_{\rm conv}$ - convective
velocity) and the chemistry.

In this letter we show that the modelling of dust formation has a
strong impact on the line shapes and the depth of the Na~I and K~I
alkali lines in dust-enshrouded L-dwarfs. We utilise our kinetic model
for dust cloud formation in atmosphere simulations of substellar
objects (brown dwarfs and extrasolar planets) in comparison to two
limiting cases in order to demonstrate the sensitive dependence of
alkali line profiles on the local $(T, p_{\rm gas}, \epsilon_i)$ profile
($\epsilon_i$ - element abundance) of the atmosphere. We show that the
treatment of dust cloud formation has a much stronger impact on the
alkali line profiles than the details of the line profile calculation.

\section[]{Atmosphere Models}

We present our studies based on simulations carried out with the
general purpose model atmosphere code {\texttt{PHOENIX}}\footnote{This
is the same code as used by Allard et al. (2001) except for the dust
chemistry and line-profile calculations.}, version 15, (Hauschildt \&
Baron 1999). We combine the study of line profiles of Johnas (2007)
with the kinetic dust cloud modelling by (Helling \& Woitke 2007)
which has recently been incorporated into {\texttt{PHOENIX}} by Dehn
(2007) (also Dehn et al. 2008).  We concentrate here on the following
two different line profile approaches for which the model atmospheres
are calculated:
\begin{itemize}
\item{\sc impact -- line profiles:}\\ 
van der Waals profiles applying the impact approximation for near line wings (Schweitzer et al. 1996)
\item{\sc modern1 -- line profiles: }\\ Detailed non-analytical,
semi-classical line profiles (Allard et al. 2005, Allard \& Spiegelman 2006, Johnas et al. 2006)
including H$_2$ in two symmetries, $C_{2v}$ and $C_{\infty v}$,
and He as perturber. The alkali line profiles are represented by two
terms, one describing the near line wing with the impact approximation
(Baranger  1958b, Royer 1971)
and the other describing the far line
wing with the one--perturber approximation in the density expansion
(Royer 1971).  \footnote{ Johnas et al. (2007a,b)  
 show that alternative alkali near line wing profiles (Mullamphy et al. 2007) 
result in negligible effects on the atmospheric structure and synthetic
spectrum.}
\end{itemize}

\noindent
For our study of the influence of different dust model approaches on
alkali line profiles, we compare simulations for the two limiting
cases {\sc Cond} and {\sc Dusty} (Allard et al. 2001), and our new
{\sc Drift} module (Dehn et al. 2008) which includes a detailed
modelling of dust formation:
\begin{itemize}
\item{{\sc Cond} approach:}\\ The {\sc Cond} approximation assumes the
dust to be in chemical and phase equilibrium with the surrounding gas
phase.  The dust is ignored as opacity source because it is {\sl
assumed} that after the dust grains have formed they sink down below
the photosphere, thus resulting in a depleted gas phase with no dust
absorption or emission. This scenario was used to model T-type brown
dwarfs (e.g. Lodieu et al. 2007) or planetary objects (e.g. Seifahrt
et al. 2007) with T$_\mathrm{eff}$ lower than $\sim$1400K but does not
allow to consider the cloud {\sl formation} process.  These models are
not realistic for the parameter range considered here and are included
for comparison only.

\item{{\sc Dusty} approach:}\\ The dust is assumed to have formed in
chemical and phase equilibrium, as in the {\sc Cond} models but is
{\sl assumed} to remain at the place of formation. In both {\sc Cond}
and {\sc Dusty}, the amount of elements bound in grains is subtracted
from the gas phase. No dust settling is taken into account. The
interstellar grain size distribution is used for the dust opacity
calculation.  {\sc Dusty} models were used at higher T$_{\rm eff}$
representing the case of a thick cloud layers inside the atmosphere of
L\,-\,dwarfs (e.g. Leggett et al. 2001) and do not consider the actual
formation process.

\item{{\sc Drift} approach:}\\ The {\sc Drift} package (Dehn et
al. 2008) is based on a kinetic treatment of dust formation (Woitke \&
Helling 2003, 2004; Helling \& Woitke 2006; Helling, Woitke \& Thi
2008).  A stationary dust formation process is assumed for application
in static model atmospheres. In these model atmospheres, seed
particles (here: TiO$_2$) nucleate from the gas phase if an
appropriate super-saturation is achieved, subsequently grow a mantle
made of various compounds (SiO$_2$[s], Al$_2$O$_3$[s], Fe[s], MgO[s],
MgSiO$_3$[s], Mg$_2$SiO$_4$[s], and TiO$_2$[s]) by 32 chemical surface
reactions (Dehn 2007), settle gravitationally, and evaporate.  The
dust formation cycle is completed by convective overshooting of
uncondensed material which is modelled by an exponential decrease of the
mass exchange frequency into the radiative zone (Ludwig et
al. 2006). The material composition of the grains, the amount and the
size of dust particles formed -- which are needed to evaluate the
cloud's opacity in the radiative transfer calculation -- are results
of our model.
\end{itemize}

\section{Results}\label{chapt:Dresults}
\begin{figure}
 \hspace*{-0.3cm}\includegraphics[width=9.7cm]{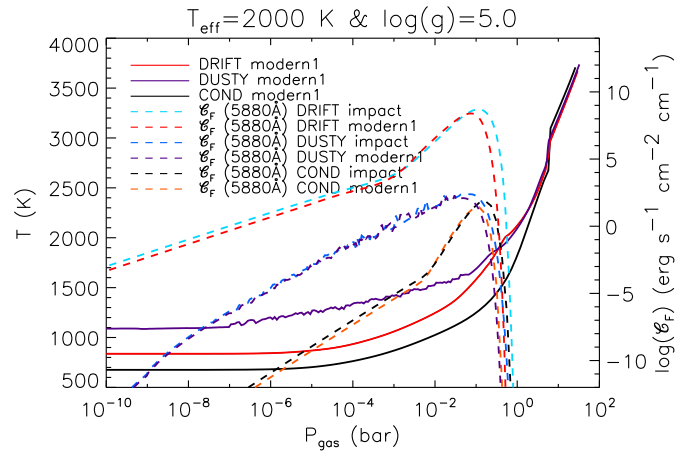}
 \caption{{\bf Left axis (solid lines)} $(T, p_{\rm gas})$ structures for the {\sc
  Drift-}{\sc Phoenix}, {\sc Dusty-}{\sc Phoenix}, and {\sc Cond-}{\sc
  Phoenix} models with the modern1 alkali line profiles for
  $T_\mathrm{eff}=2000$K, $\log(g) =5.0$.  {\bf Right axis (dashed lines):} Flux
  contribution function at 5880\,\AA\,of the six different model ({\sc
  Drift}, {\sc Dusty}, and {\sc Cond} models with {\sc impact} and
  {\sc modern1} line profiles).}
\label{fig:dT-p}
\end{figure}
\begin{figure}
 \hspace*{-0.3cm}\includegraphics[width=9.5cm]{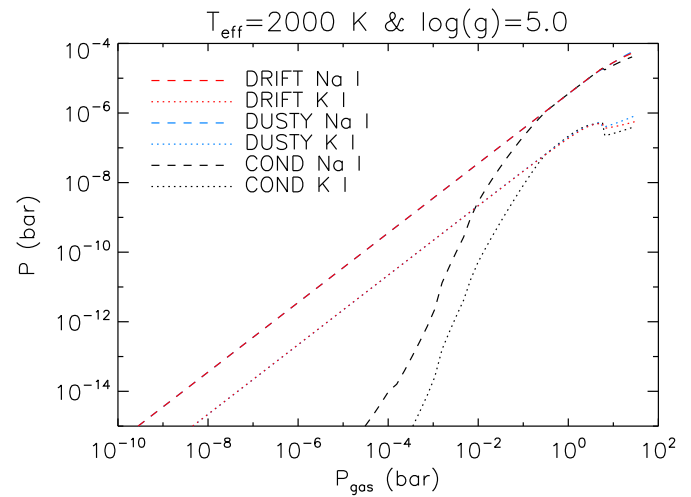}
 \caption{The partial pressure p [bar] of Na and K as function of gas pressure p$_{\rm gas}$ [bar] as results of the model runs for the {\sc Drift}, {\sc Dusty}, and {\sc Cond} dust model approaches.}
\label{fig:pNaK}
\end{figure}
\begin{figure}[t]
\hspace*{-0.3cm}\includegraphics[width=8.9cm]{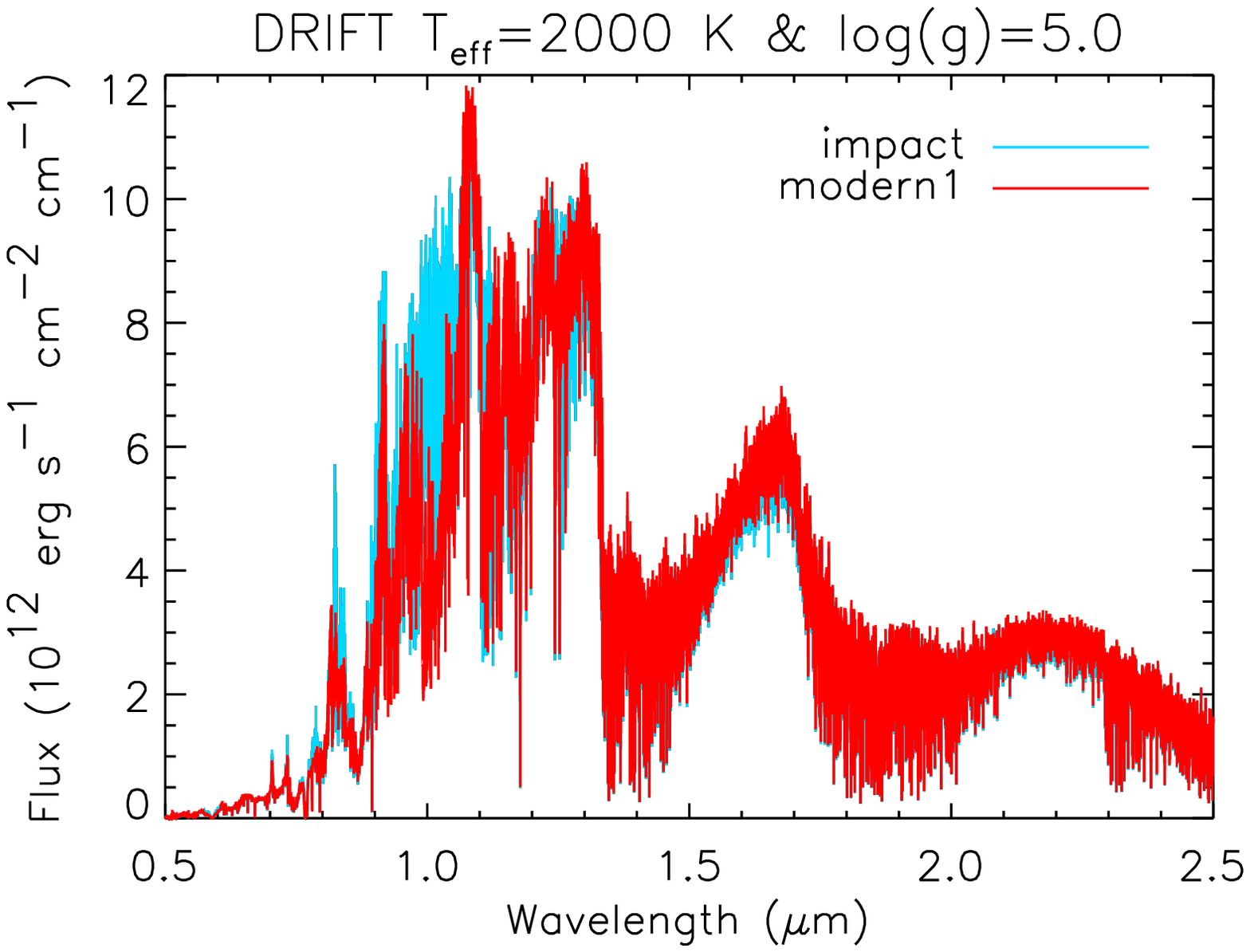}\\
\hspace*{-0.3cm}\includegraphics[width=8.9cm]{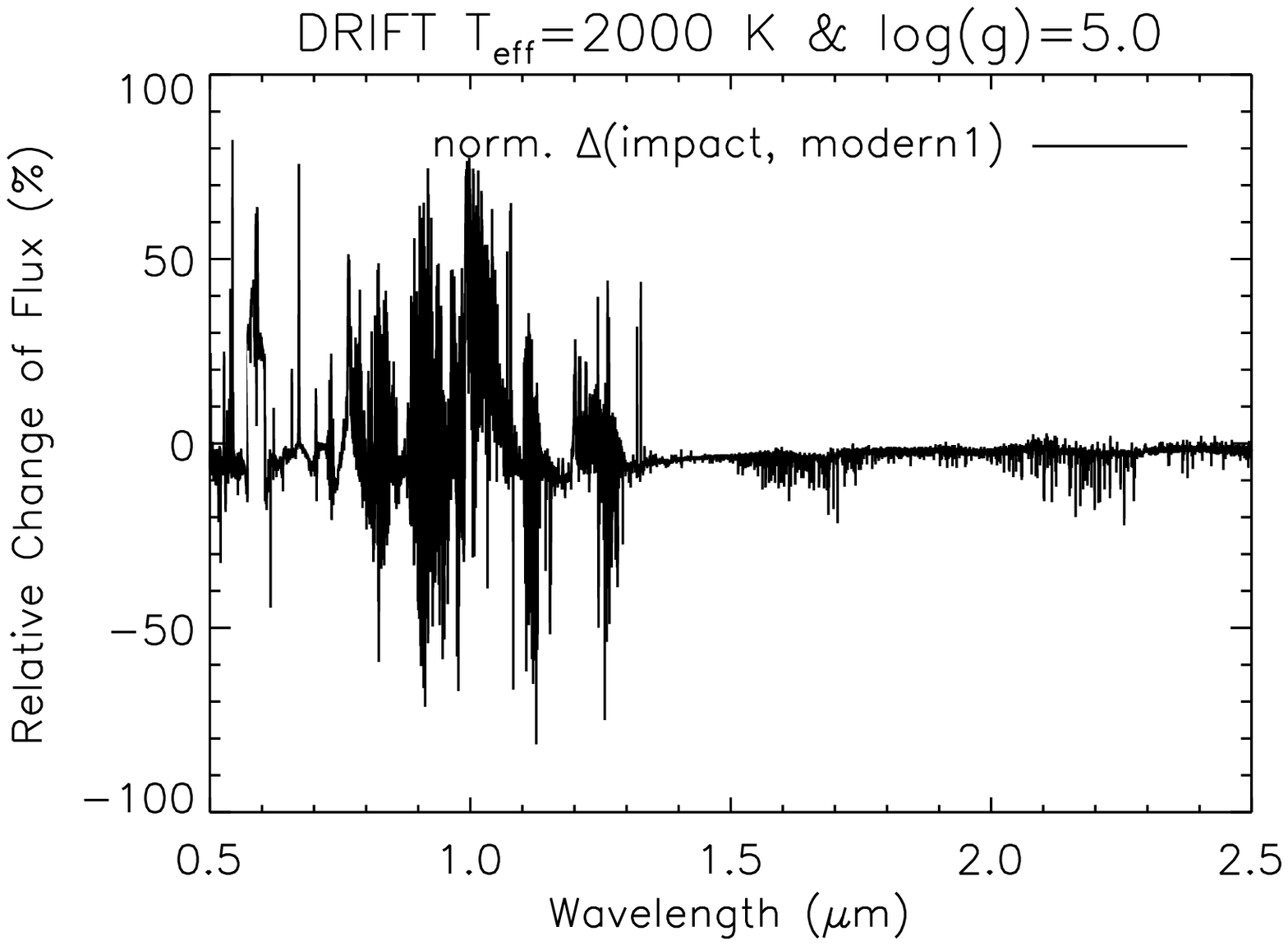}
\caption{{\bf Top:} {\sc Drift-Phoenix} model atmospheres for T$_{\rm eff}=2000$K, $\log$ g =5. computed with the {\sc impact}
and {\sc modern1} line profiles. {\bf Bottom:}  Relative changes of the top panel fluxes  $\Delta(A, B)=(A-B)/A$.}\label{fig:DRIFTmodimp}
\end{figure}
\begin{figure}[t]
\hspace*{-0.3cm}\includegraphics[width=8.9cm]{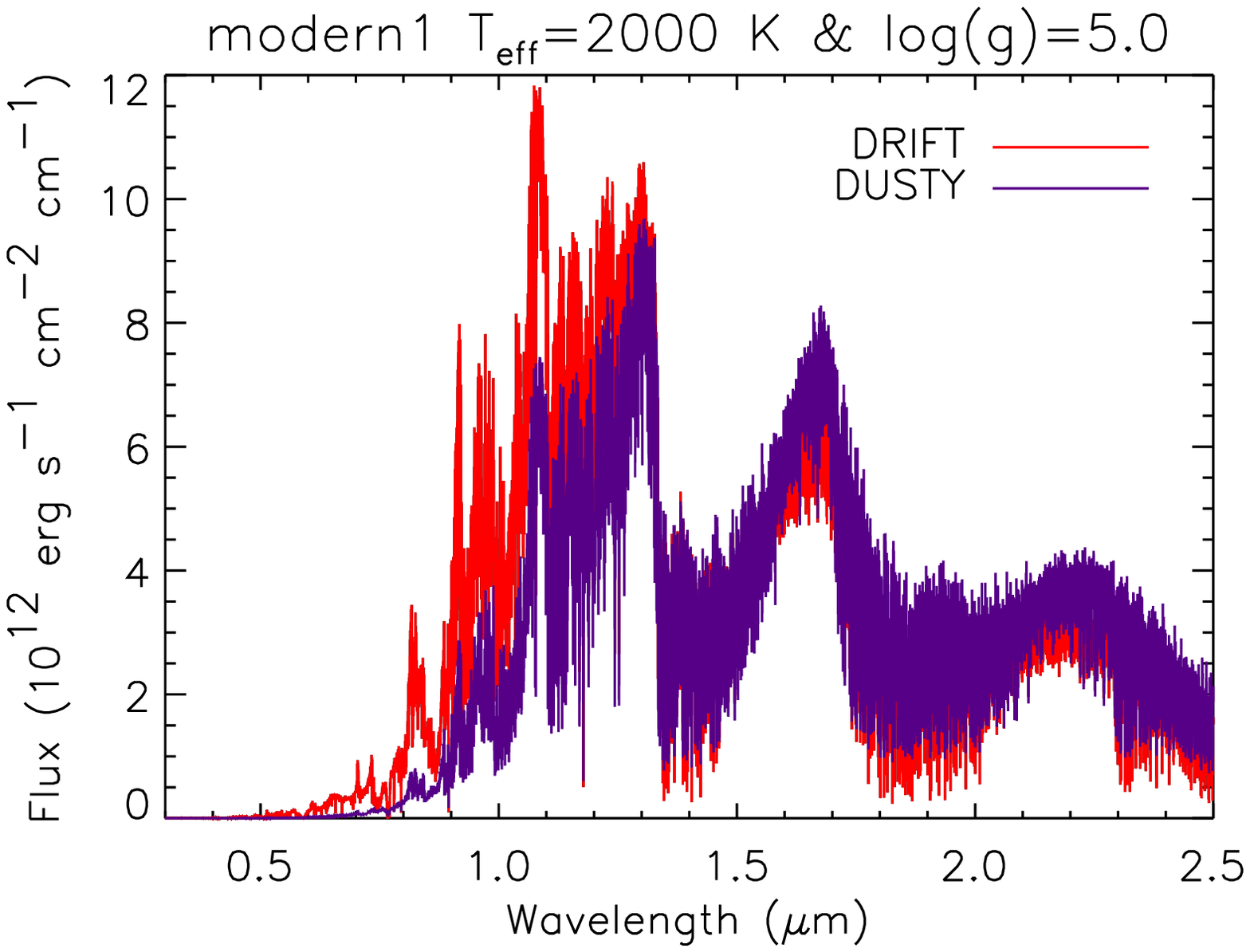}\\
\hspace*{-0.3cm}\includegraphics[width=8.9cm]{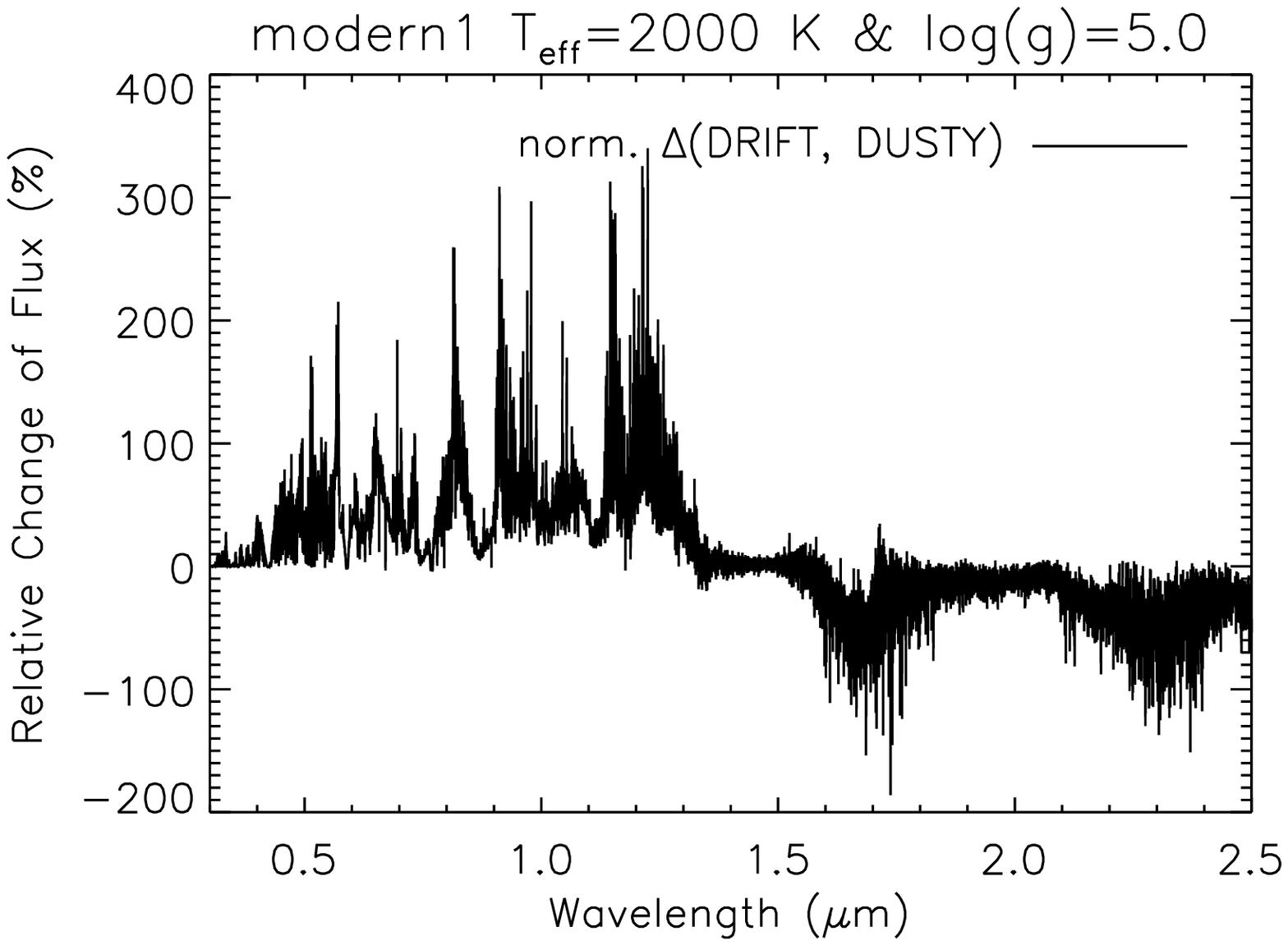}
\caption{{\bf Top:} {\sc Dusty-Phoenix}  and {\sc Drift-Phoenix}  model atmospheres for T$_{\rm eff}=2000$K, $\log$ g =5 computed with the modern1 -- line profiles.  {\bf Bottom:} Relative  change  of the top panel fluxes   $\Delta(A, B)=(A-B)/A$.}\label{fig:mod1DD}
\end{figure}


We demonstrate our results for a L-type brown dwarf with
$T_\mathrm{eff}=2000$\,K, $\log(g)=5.0$, and solar composition (Anders
\& Grevesse 1986).  Figure~\ref{fig:dT-p} shows that the atmospheric
$(T, p)$--structures are very similar for $p_{\rm gas}> 1$\,bar for {\sc
Drift-Phoenix} and {\sc Dusty-Phoenix} models, but can differ by up to
300K (i.e. 30\,\%) at lower pressures. All models agree well in the
innermost atmospheric layers.  The {\sc Drift-Phoenix} temperature
gradient is steeper compared to the {\sc Dusty-Phoenix} case. The {\sc
Drift-Phoenix} model structure falls basically between the two
limiting cases {\sc Dusty} and {\sc Cond}. The different line profiles
cause negligible differences in the $(T, p)$--structures for all three
cloud models (Johnas 2007).

\begin{figure*}
\hspace*{-0.5cm}\includegraphics[width=9.1cm]{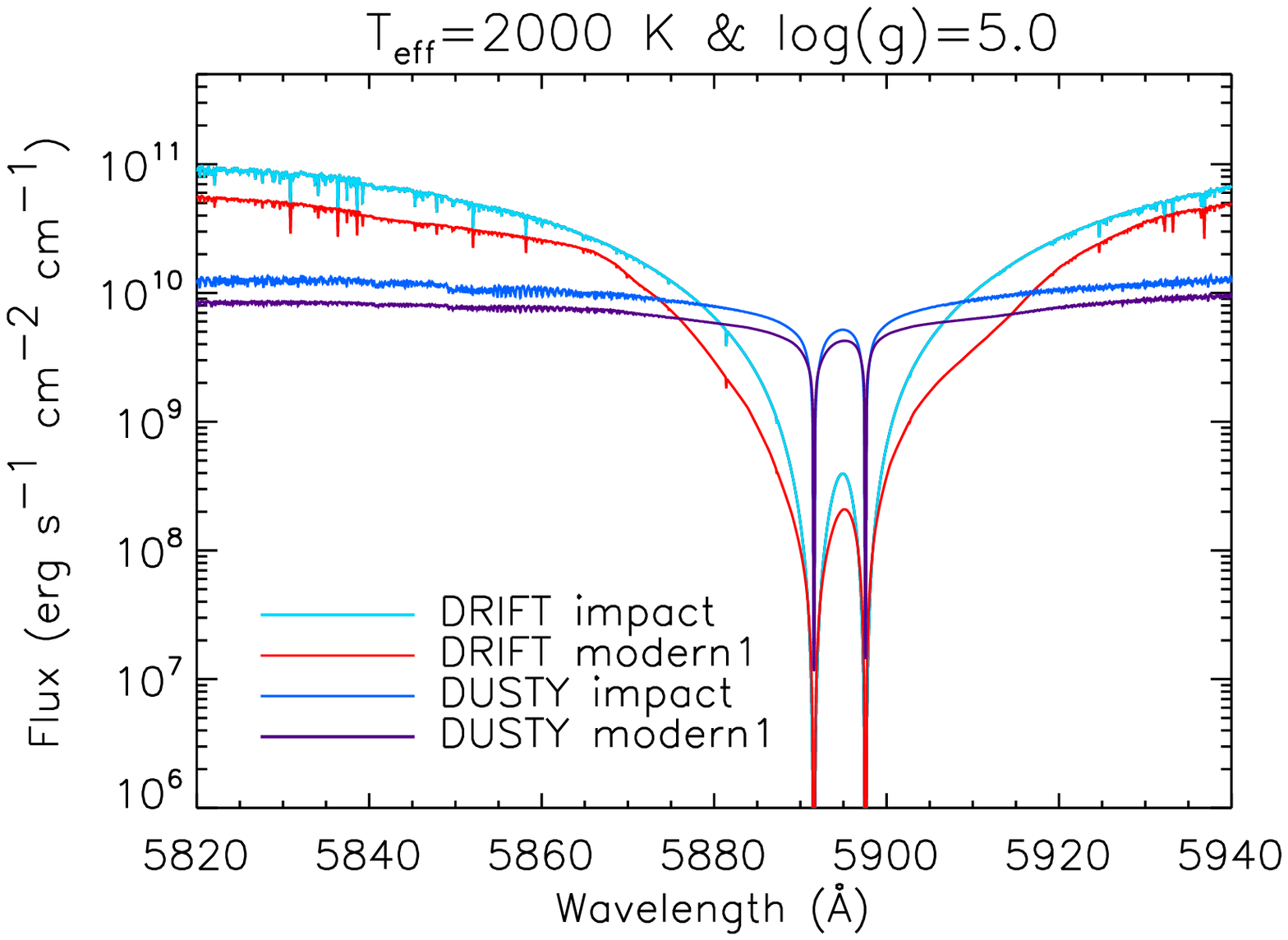}
\hspace*{-0.3cm}\includegraphics[width=9.1cm]{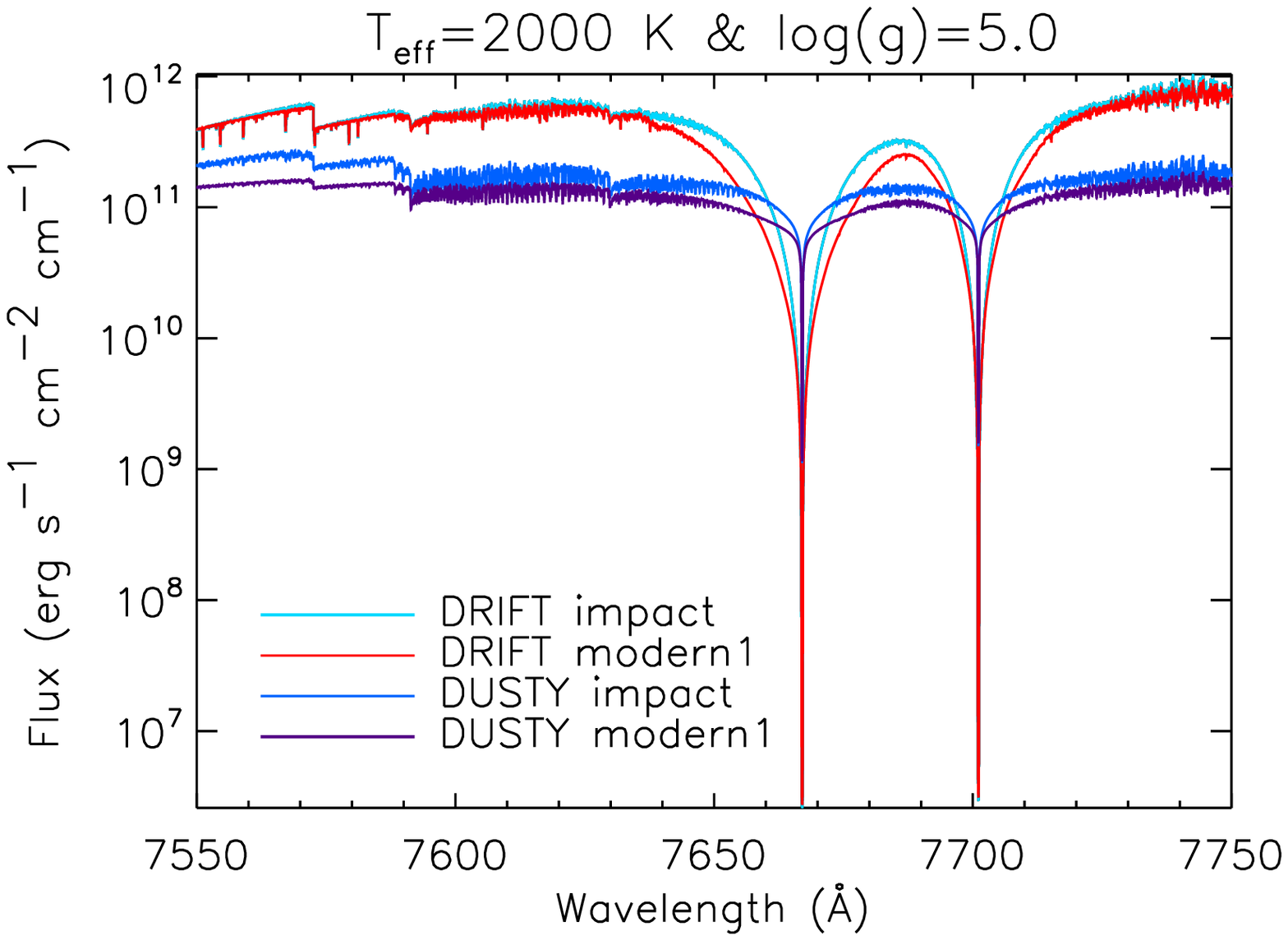}
\caption{Synthetic spectra displaying line profiles of the alkali species with the
highest concentration in cool atmospheres, the Na\,I\,D ({\bf left}) and
K\,I ({\bf right}) doublets, computed with {\sc Drift-Phoenix} and {\sc Dusty-Phoenix}  models with
impact -- and modern1 -- line profiles.}
\label{fig:dNaK}
\end{figure*}

In addition to the $(T, p)$--structures in Fig.~\ref{fig:dT-p}, the
flux contribution function $\mathscr{C}_F$ (see Fuhrmeister et
al. 2006) of each of the simulations is shown at 5880\,\AA\ which is a
line wing wavelength in the Na\,I~D$_2$ resonance line
(Fig.~\ref{fig:dNaK}).  The maximum of $\mathscr{C}_F$ is correlated
to the location in the atmosphere at which the line wing is formed at
that specific wavelength (Magain 1986, Fuhrmeister et al. 2006).
Note, however, that $\mathscr{C}_F$ should only be considered for
strong lines like Na\,I and K\,I.  Figure~\ref{fig:dT-p} demonstrates
that in general the Na\,I\,D$_2$ line wings at 5880\,\AA\, form at
comparable local gas pressures in the model atmospheres despite the
different dust cloud approaches considered here. We furthermore find
that Na and K appear with comparable concentrations in the line
forming regime in all models (Fig.~\ref{fig:pNaK}). {\it This leaves the local
temperature as only candidate for potential differences in the Na~I and K~I
line profile for different dust cloud treatment.}

Figure~\ref{fig:DRIFTmodimp} shows that the low resolution {\sc
Drift-Phoenix} spectrum for two different line profile setups will
basically differ in the optical and near-IR region which is, hence,
the spectral region predominantly influenced by the line profile
treatment. Although one might have expected more effects in larger
spectral ranges, the relative changes shown in the bottom-panel of
Fig.~\ref{fig:DRIFTmodimp} are relatively small. On the contrary, the
differences in the low-resolution spectra between the different dust
approaches ({\sc Drift}, {\sc Dusty}) for fixed line profile treatment
are quite remarkable (Fig.~\ref{fig:mod1DD}).  These strong
differences are no surprise because the atmospheric $(T,
p)$--structures differ substantially (Fig.~\ref{fig:dT-p}), and
therefore the amount of molecular opacity carriers does change. For
example, {\sc Dusty-Phoenix} produces a TiO-concentration of two orders
of magnitude less than {\sc Drift-Phoenix} at $p_{\rm
gas}\approx\,10^{-4}$\,bar. The temperature-dependent gas-phase
composition does also influence the dust formation by providing the
constituents from which the dust forms. We find similar discrepancies
when comparing spectra between {\sc Drift-Phoenix} and {\sc
Cond-Phoenix} models (not shown).

\begin{figure*}
\hspace*{-0.3cm}
 \includegraphics[width=19cm]{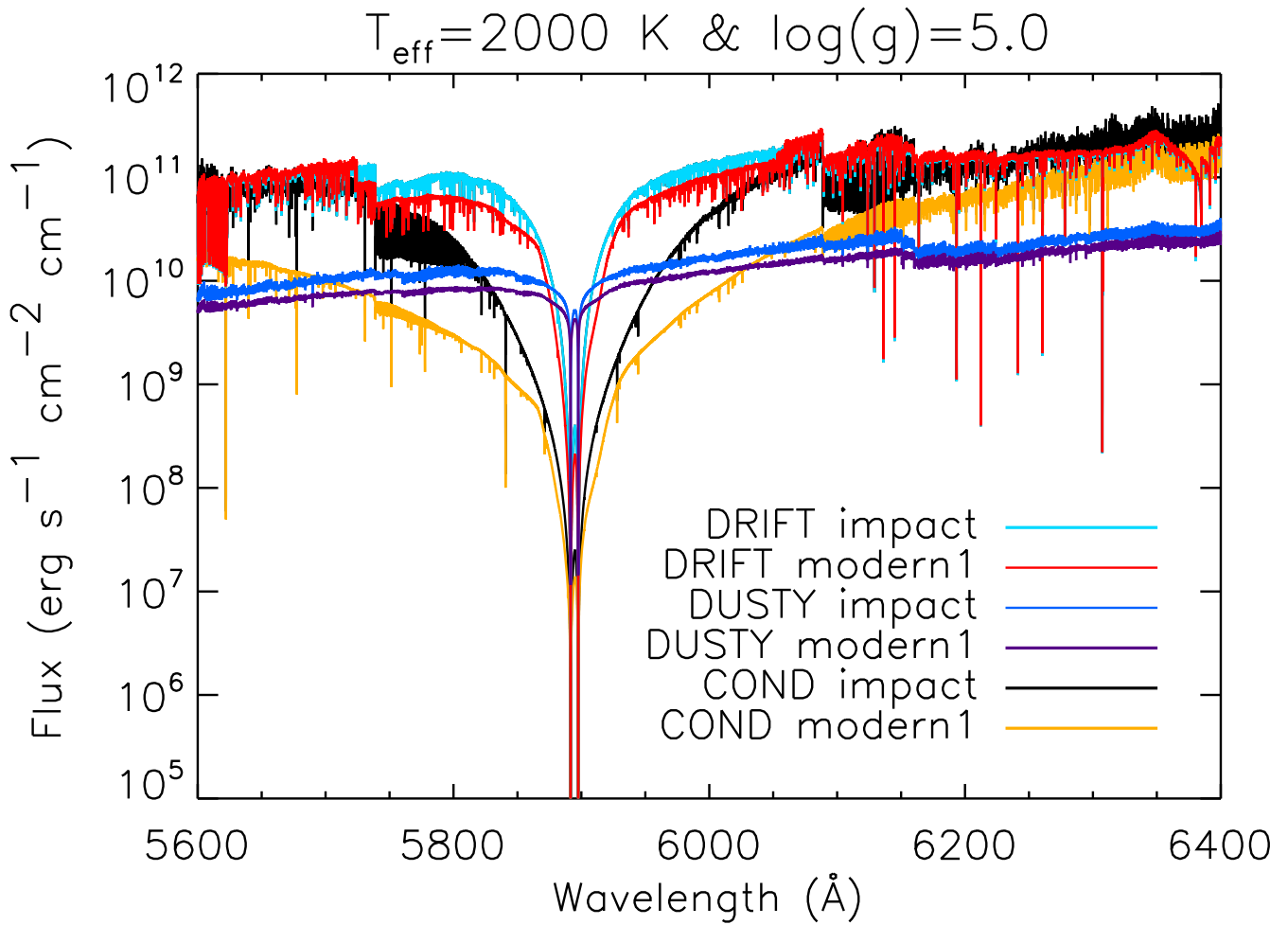}
 \caption{The Na\,I\,D doublet as computed with the {\sc impact-} and {\sc modern1-}line profile
 setups in  {\sc Drift-Phoenix}, {\sc Dusty-Phoenix}, and {\sc Cond-Phoenix} simulations.} \label{fig:COND}
\end{figure*}

\subsection{Alkali lines}\label{ss:al}
The classical {\sc Dusty-Phoenix} simulation result in a lower
pseudo--continuum for both, the Na\,I\,D and K\,I doublets compared to
the {\sc Drift-Phoenix} results (Fig.~\ref{fig:dNaK}).  The {\sc
Dusty-Phoenix} models have a higher dust opacity than the {\sc
Drift-Phoenix} models which results in strongly differing $(T,
p)$-structures. Hence, also the line shapes for the Na\,I\,D and K\,I
doublets differ considerably. It appears that the local gas pressure
and the concentration of Na~I and K~I are very similar in the line
forming region in these models but large differences occur in the
local temperatures. Comparing Fig.~\ref{fig:COND} visualises that the
{\sc Cond-Phoenix} line profiles match up with the Drift results
better in the far wings since here the differences in the local
temperature are somewhat smaller (Fig.~\ref{fig:dT-p}). Generally, the
{\sc Drift-Phoenix} atmosphere simulations, which are based on a detailed
kinetic cloud model, yield a stronger pseudo-continuum and deeper
line cores. The similarity of the alkali partial pressures especially
between the {\sc Drift} and {\sc Dusty} approach point to the local temperature as
main cause for the large differences in the line-profiles. All figures
show that {\it the observable line shapes depend more on the dust model  than on
the line profile model.}

\section{Conclusion}\label{chapt:con}

We have demonstrated that the dust treatment in an atmosphere simulation for
substellar objects has a large influence on the resulting line-profile
of the alkalies Na~I and K~I.  We conclude that the main cause for
these large differences is the strongly changing local temperature
which is directly linked to the dust treatment by the resulting dust
opacity. The different treatments of the alkali line profiles result
in negligible changes compared to the effect of dust treatment.

The pseudo-continuum of alkali line-profiles are strongest if based on
a microphysical dust model taking into account the kinetic nature of
dust formation ({\sc Drift-Phoenix}). This findings may foster studies
on Li~I and Rb~I lines since Johnas (2007) has shown that in
particular the far K~I line wings can mask the Rb~I doublet, but both
the Na~I and the K~I overlap the Li~I doublet cores.

\section*{Acknowledgments}

We thank the referee for valuable commends on the manuscript.
This work was supposed in part by the DFG via Graduiertenkolleg 1351.
Some of the calculations presented here were performed at the H\"ochstleistungs
Rechenzentrum Nord (HLRN); at the NASA's Advanced Supercomputing Division's
Project Columbia, at the Hamburger Sternwarte Apple G5 and Delta Opteron
clusters financially supported by the DFG and the State of Hamburg; and at the
National Energy Research Supercomputer Center (NERSC), which is supported by
the Office of Science of the U.S.  Department of Energy under Contract No.
DE-AC03-76SF00098.  We thank all these institutions for a generous allocation
of computer time.

\end{document}